\documentclass{webofc}
\usepackage[varg]{txfonts}   
\newlength{\colw}
\graphicspath{{./figures/}}

\providecommand{\half}{\frac{1}{2}}

\newcommand{\order}{{\cal O}}
\newcommand{\One}{1\kern-4.5pt1}

\newcommand{\om}{\omega}

\begin{document}
\title{Hadrons at high temperature}
%
%
\subtitle{An update from the FASTSUM collaboration}

\author{\firstname{Jon-Ivar} \lastname{Skullerud}\inst{1,3}\fnsep\thanks{\email{jonivar.skullerud@mu.ie}} \and
  \firstname{Gert} \lastname{Aarts}\inst{2}\and
  \firstname{Chris} \lastname{Allton}\inst{2}\and
  \firstname{M. Naeem} \lastname{Anwar}\inst{2}\and
  \firstname{Ryan} \lastname{Bignell}\inst{2}\and
  \firstname{Tim} \lastname{Burns}\inst{2}\and
  \firstname{Sergio} \lastname{Chaves García-Mascaraque}\inst{2} \and
  \firstname{Simon} \lastname{Hands}\inst{3}\and
  \firstname{Rachel} \lastname{Horohan D'Arcy}\inst{1}\and
  \firstname{Benjamin} \lastname{J\"ager}\inst{4}\and
  \firstname{Seyong} \lastname{Kim}\inst{5}\and
  \firstname{Maria Paola} \lastname{Lombardo}\inst{6}\and
  \firstname{Eoghan} \lastname{Murphy}\inst{1}\and
  \firstname{Sam} \lastname{Offler}\inst{2}\and
  \firstname{Ben} \lastname{Page}\inst{2}\and
  \firstname{Sin\'ead M.} \lastname{Ryan}\inst{7}\and
  \firstname{Thomas} \lastname{Spriggs}\inst{2}\and
  \firstname{Dawid} \lastname{Stasiak}\inst{2}\and
  \firstname{Felix P.G.} \lastname{Ziegler}\inst{8}
}

\institute{Department of Theoretical Physics, National University of
  Ireland Maynooth, Maynooth, Co Kildare, Ireland 
\and
Department of Physics, Swansea University, Swansea, SA2 8PP, United Kingdom
\and
Department of Mathematical Sciences, University of
Liverpool, Liverpool L69 3BX, United Kingdom
\and
CP3-Origins \& Danish IAS, Department of Mathematics and
Computer Science, University of Southern Denmark, Campusvej 55, 5230 Odense M,
Denmark
\and
Department of Physics, Sejong University, Seoul 05006, Korea
\and
INFN, Sezione di Firenze, 50019 Sesto Fiorentino (FI), Italy
\and
School of Mathematics and Hamilton Mathematics Institute, Trinity College Dublin, Ireland
\and
School of Physics and Astronomy, The University of Edinburgh, Edinburgh EH9 3FD, United Kingdom
}

\abstract{%
We present the most recent results from the FASTSUM collaboration for hadron properties at high temperature. This includes the temperature dependence of the light and charmed meson and baryon spectrum, as well as properties of heavy quarkonia. The results are obtained using anisotropic lattices with a fixed scale approach. We also present the status of our next generation gauge ensembles.
}
\maketitle
\section{Introduction}
\label{intro}

The fate of hadrons under extreme conditions is one of the outstanding questions in the theory of strong interactions, quantum chromodynamics (QCD). 
As the temperature increases, the hadron gas --- with confined quarks and broken chiral symmetry --- transitions into a quark--gluon plasma (QGP), with deconfined light degrees of freedom and chiral symmetry restored.  In the vicinity of the crossover transition, thermal modifications of hadron properties are expected as a precursor to chiral symmetry restoration and deconfinement.  Hadrons built out of heavier (charm and beauty) quarks may survive in the QGP, and these states may serve as probes of the hot medium created in heavy-ion collisions.

Hadron properties are encoded in the spectral functions $\rho_\Gamma(\om)$ (where $\Gamma$ denotes the quantum numbers), which are related to the euclidean correlators $G(\tau)$ computed on the lattice by an integral transform,
\begin{equation}
G(\tau;T) = \frac{1}{2\pi}\int_0^\infty\rho(\om;T)K(\tau;T)d\om\,,
\label{eq:spectral}
\end{equation}
where $K(\tau;T)$ is a known kernel that depends on the nature of the correlator.  Determining $\rho(\om)$ from $G(\tau)$ is an ill-posed problem, and no single method has to date been shown to yield reliable results in all circumstances.  Where possible, avoiding this inversion problem by extracting information on thermal modifications directly from the correlator may be preferable.  A useful tool in this context is the \emph{reconstructed correlator}, which is obtained by integrating eq.~\eqref{eq:spectral} with $\rho(\om;T)$ at some reference temperature $T=T_r$, which is usually taken to be the lowest available temperature.  In many cases this may be constructed directly from the correlator $G(\tau;T_r)$ \cite{Ding:2012sp} without having to determine $\rho(\om;T_r)$.

The FASTSUM collaboration has employed anisotropic lattice QCD to study properties of charmonium and open-charm hadrons \cite{Aarts:2007pk,Kelly:2018hsi}, beautonium \cite{Aarts:2010ek,Aarts:2011sm,Aarts:2013kaa,Aarts:2014cda} and light and strange baryons \cite{Aarts:2015mma,Aarts:2017rrl,Aarts:2018glk}, as well as the electrical conductivity of QCD matter \cite{Amato:2013naa,Aarts:2014nba} and properties of the chiral transition \cite{Aarts:2014nba,Aarts:2020vyb}.  In these proceedings we will present recent results for mesons and baryons containing light, strange, charm and beauty quarks using our most recent gauge ensembles, and report on the status of our next generation ensembles.

\section{Simulation details}
\label{sec:lat-params}

\begin{table}[ht]
\centering
\begin{tabular}{c|crrcrrr}\hline
Gen & $N_f$& $\xi$  & $a_s$ (fm) & $a_\tau^{-1}$ (GeV) & $m_\pi$ (MeV) & $N_s$
& $L_s$ (fm)  \\ \hline
1 \cite{Aarts:2007pk} & 2   & 6.0 &  0.162 &   7.35 & 490 & 12 & 1.94 \\
2 \cite{Aarts:2014nba} & 2+1 & 3.45 &  0.121 &   5.63 & 390 & 24 & 2.95 \\
 & & & & & & 32 & 3.94 \\
2L \cite{Aarts:2020vyb} & 2+1 & 3.45 &  0.112 &   6.08 & 240 & 32 & 3.58 \\\hline
2P & 2+1 & 3.45 & *0.100 &  *6.80 & 140 & 48 & 4.80 \\
3 & 2+1 & 7.0 & *0.120 & *11.66 & *390 & 32 & 3.94 \\
\hline
\end{tabular}
\caption{FASTSUM ensemble parameters: number of flavours $N_f$, spatial lattice spacing $a_s$, inverse temporal lattice spacing $a_\tau^{-1}$, anisotropy $\xi=a_s/a_\tau$, pion mass $m_\pi$, number of lattice sites $L_s$ in the spatial directions and spatial extent $L_s$.  The Gen2P and Gen3 ensembles are in preparation (see section~\ref{sec:tuning}) and the starred numbers are target or expected values.}
\label{tab:params}
\end{table}

Our simulations are carried out using anisotropic lattices with an
$\order(a^2)$ improved gauge action and an $\order(a)$ improved Wilson
fermion action with stout smearing.  The Gen2 and Gen2L ensembles were produced following the parameter tuning and ensembles generated by the Hadron Spectrum Collaboration \cite{Edwards:2008ja,Lin:2008pr}.  Details about all our ensembles
used to date as well as our planned new ensembles (Gen2P and Gen3) are
given in table~\ref{tab:params}.  The temperature is given by $T=(a_\tau N_\tau)^{-1}$ and is varied by changing the number of sites $N_\tau$ in the temporal direction.  For more details about the Gen2 and Gen2L ensembles, see Ref.~\cite{Aarts:2020vyb} and references therein. Most of the results presented in
this update are from our most recent ensemble, Gen2L.

\section{Results}
\label{sec:results}

\subsection{Light hadrons}
\label{sec:light}

We have previously \cite{Aarts:2015mma,Aarts:2017rrl,Aarts:2018glk}
investigated light and strange baryon
correlators below and above the chiral crossover.  We have found that
in all channels studied, the positive-parity baryon masses remain
constant to within the accuracy of our calculations below $T_c$, while
the negative-parity baryon masses decrease with increasing temperature
and become degenerate with the positive-parity masses at $T_c$.  The
parity doubling in the correlators can be encoded in the $R$-parameter
defined by
\begin{align}
R &= \frac{\sum\limits_{n=n_0}^{N_\tau/2-1}R(\tau_n)/\sigma^2(\tau_n)}
{\sum\limits_{n=n_0}^{N_\tau/2-1}1/\sigma^2(\tau_n)}\qquad\text{with}\qquad&
R(\tau) &= \frac{G(\tau)-G(N_\tau-\tau)}{G(\tau)+G(N_\tau-\tau)}\,\,
\label{eq:baryon-R}
\end{align}
where $\sigma(\tau)$ is the uncertainty in the correlator $G(\tau)$ and we choose $n_0\sim4$ to reduce contamination from lattice artefacts and excited states.
We have found that $R$ behaves qualitatively like an order parameter
for the chiral transition, and that the inflection point in $R$ is
consistent between different baryon (spin and flavour) channels and
with $T_c$ determined from the renormalised chiral condensate and
chiral susceptibility.  These studies were carried out using standard
exponential fits with extended (smeared) sources to enhance the
overlap with the ground state, but the results have also been
corroborated by extracting spectral functions with the maximum entropy
method (MEM).

\begin{figure}[t]
\centering
\includegraphics[width=\colw]{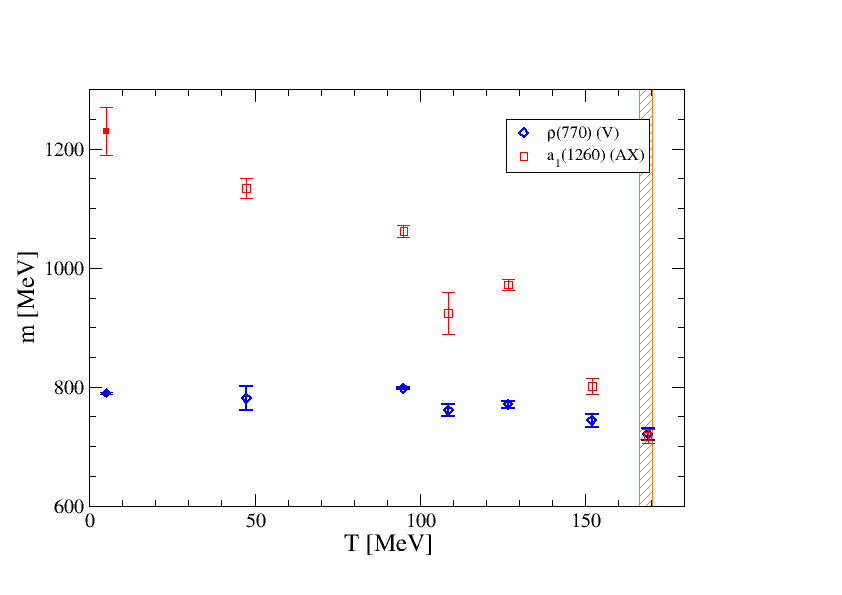}
\includegraphics[width=\colw]{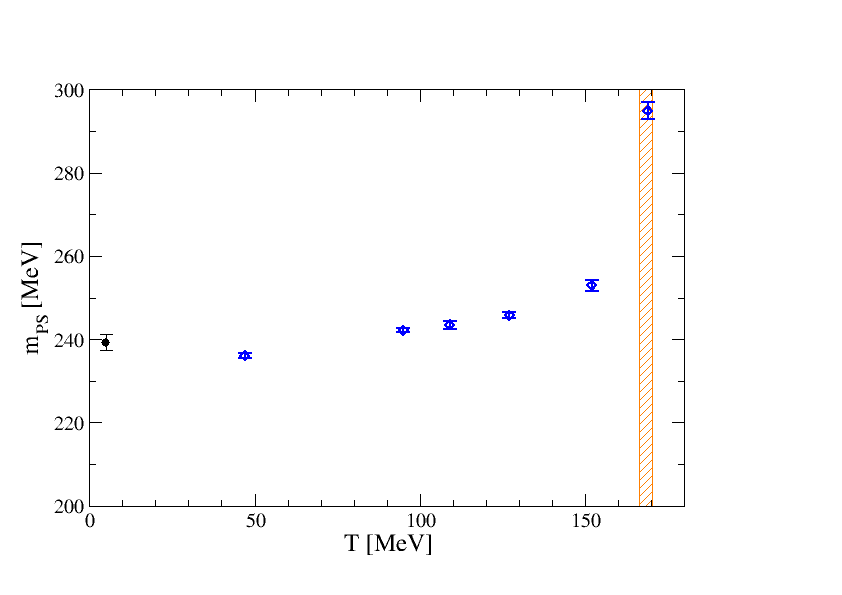}
\caption{Light meson masses as function of temperature, from analysis
  of smeared correlators. Left: vector
  and axial-vector mesons, $\rho$ and $a_1$.  Right: pion.
  The vertical band indicates the chiral crossover.  The solid point for the $\rho$ at $T\approx0$ represent the mass determined by the Hadron Spectrum collaboration on the same ensemble in a coupled-channel calculation using distillation and a variational analysis \cite{Wilson:2015dqa,Wilson:2019wfr}, while that for $a_1$ is the Particle Data Group value for the mass.}
\label{fig:light}
\end{figure}

We have recently applied the methods used for studying light baryons
to light, strange and charmed mesons as well as charmed baryons.
Specifically, we have performed exponential fits to correlators of
extended operators with optimised overlaps with the ground state, to
determine hadron masses below and in the vicinity of $T_c$.
Fig.~\ref{fig:light} shows the result of this analysis for light
mesons in the vector ($\rho$), axial-vector [$a_1(1260)$] and pseudoscalar
(pion) channels.  The vector and axial-vector are chiral partners and
should be degenerate if chiral symmetry is restored.  As we can see in
fig.~\ref{fig:light} (left) this near-degeneracy\footnote{Since the light quark mass is nonzero we do not expect exact degeneracy.} becomes manifest at $T_c$
with the mass of the axial-vector meson dropping substantially to
become approximately degenerate with the $\rho$ meson.  It is interesting to note that the
$\rho$ meson also appears to exhibit a small negative mass shift below
$T_c$.  A similar effect is seen for the kaon.

In the right panel of fig.~\ref{fig:light} we show the pion mass as a function of temperature.  We see that unlike the $\rho$, the pion mass increases with temperature.  This can be understood as a precursor to chiral symmetry restoration, as the pion is the pseudo-Goldstone boson of the broken chiral symmetry and as a result of this is much lighter than the other hadrons below $T_c$.  This mass difference will disappear once chiral symmetry is restored.  At this point, the pion (pseudoscalar) and scalar meson (isovector $a_0$) should become degenerate; however, it is numerically difficult to reliably extract masses of scalar mesons and we have as yet not been able to observe this degeneracy.

\subsection{Charmed hadrons}
\label{sec:charm}
\begin{figure}[h]
\centering
\includegraphics[width=\colw]{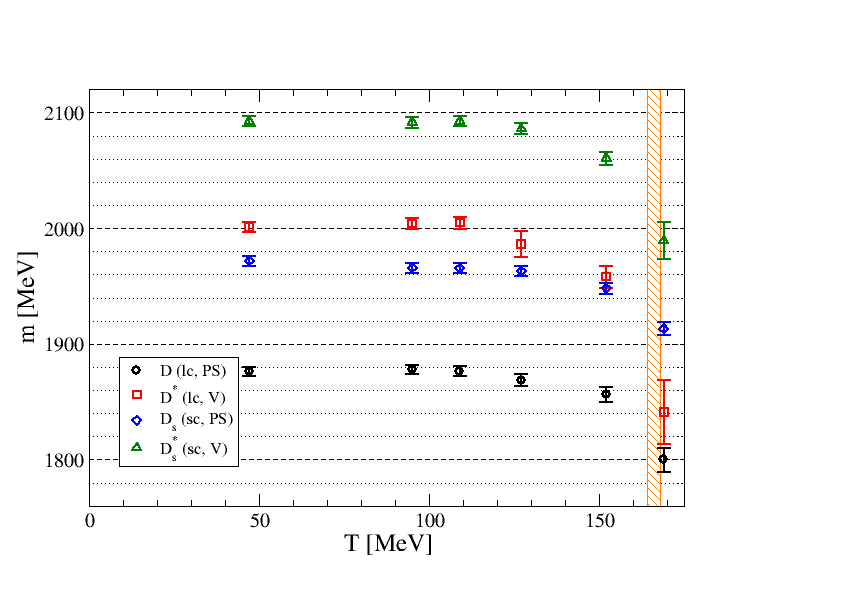}
\includegraphics[width=\colw]{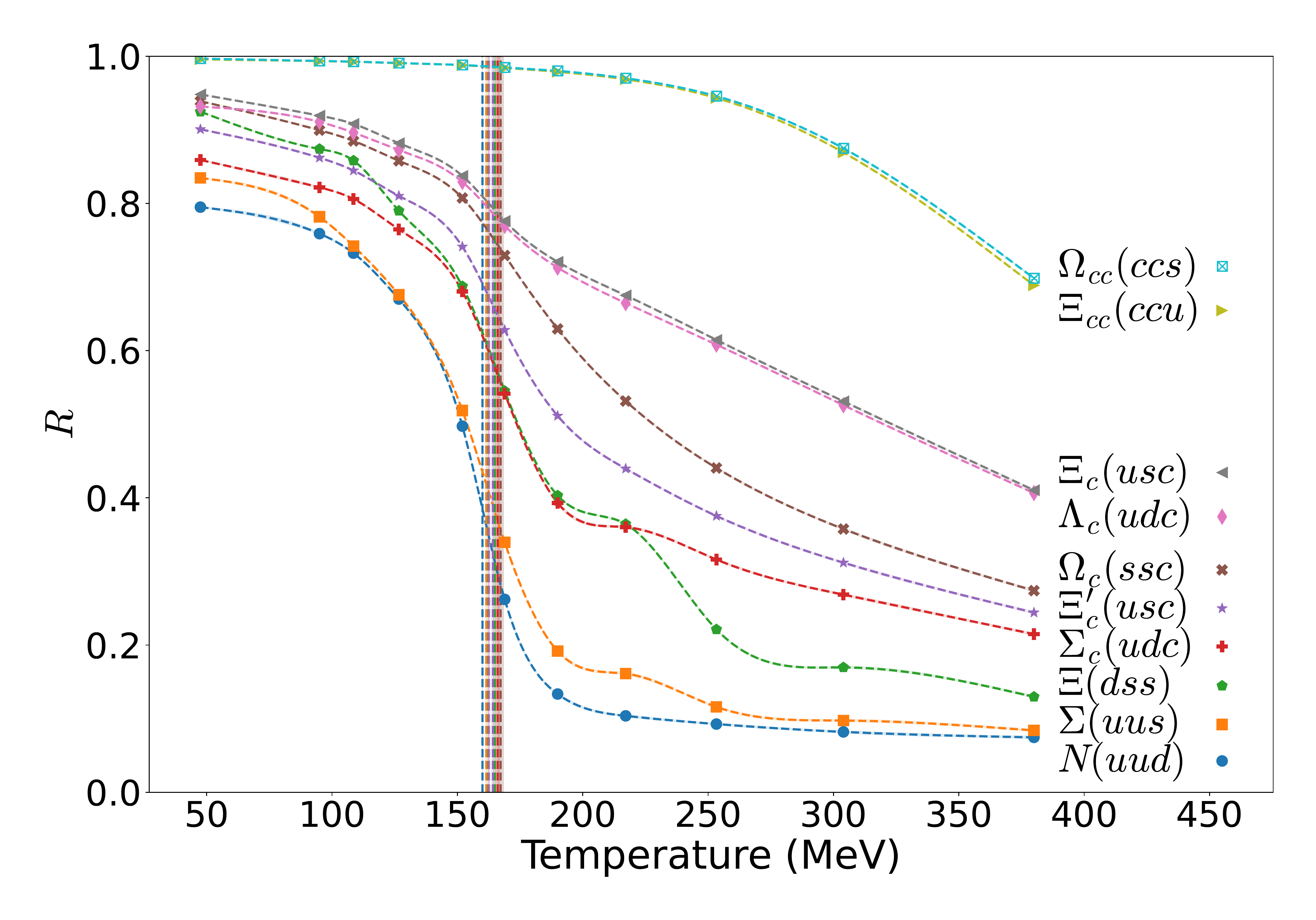}
\caption{Left: open charm masses from analysis of smeared correlators \cite{Aarts:2022krz}.
  Right: the parity doubling parameter $R$ for spin-$\frac{1}{2}$
  charmed baryons \cite{Bignell:InPrep}. The vertical band indicates the chiral crossover.}
\label{fig:opencharm}
\end{figure}

Temporal correlators and spectral functions of open-charm mesons were studied for the first time by us using the Gen2 ensemble \cite{Kelly:2018hsi,Quinn:2019uwq}.  Recently, this has been updated with a detailed study \cite{Aarts:2022krz} of mass modifications below and around $T_c$ using our Gen2L ensemble.  In the left panel of fig.~\ref{fig:opencharm} we show the resulting masses of the ground state open-charm mesons in the pseudoscalar ($D,D_s$) and vector ($D^*,D_s^*$) channels.  We find no change for $T\lesssim120\,$MeV, but a decrease of about 40\,MeV in all channels at $T\sim150\,$MeV.  We find somewhat larger modifications in the vector than the pseudoscalar channel, and an indication that the strange $D_s$ meson is less affected than the others.  This may have implications for the interpretation of experimental results suggesting an increased $D_s/D$ ratio in $A-A$ vs $p-p$ collisions.

We have also investigated charmed baryons \cite{Bignell:InPrep}.  The right panel of fig.~\ref{fig:opencharm} shows the parity-doubling $R$ parameter \eqref{eq:baryon-R} for light, singly charmed and doubly charmed spin-$\half$ baryons.  We see that the singly-charmed baryons exhibit the same qualitative behaviour as light baryons, even to the extent that the inflection point is consistent.  For doubly charmed baryons on the other hand we do not find any clear signal of a transition.  Similar results are found for spin-$\frac{3}{2}$ baryons, but in this case also some of the doubly-charmed baryons exhibit features consistent with a transition.  We are in the process of analysing these results in terms of thermal mass modifications.

\begin{figure}[h]
\centering
\includegraphics[width=\colw]{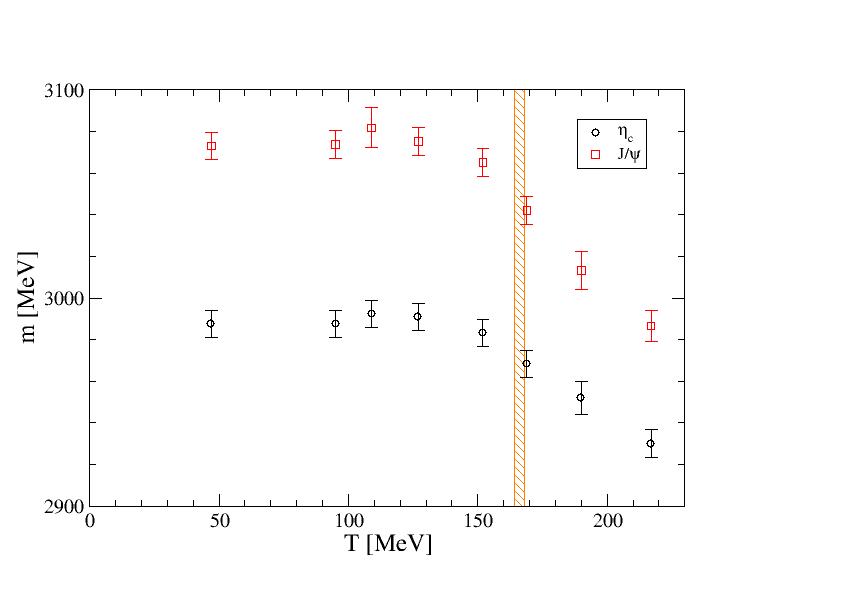}
\includegraphics[width=\colw]{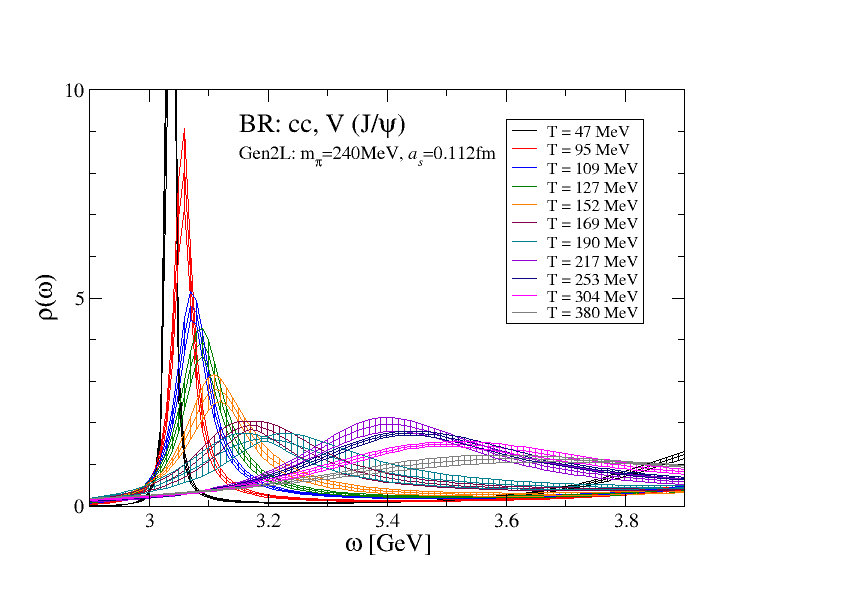}
\caption{Left: 
  $J/\psi$ and $\eta_c$ masses from analysis of smeared correlators.
  The vertical band indicates the chiral crossover.  As the analysis assumes narrow bound states, the results for $T>T_c$ cannot be taken as evidence that these states survive in the quark--gluon plasma. Right:
  $J/\psi$ spectral functions from the BR method.}
\label{fig:charmonium}
\end{figure}

In fig.~\ref{fig:charmonium} we show preliminary results for charmonium from our Gen2L ensemble using two different analysis methods.  The left panel shows masses determined with the same approach as already detailed for light and open-charm hadrons.  We see indications of a very slight negative mass shift just below $T_c$.  Whether the $J/\psi$ and $\eta_c$ survive in the QGP and, if so, to what temperature, remains unresolved, and the results shown for $T>T_c$ should not be taken as evidence of survival as they have been obtained assuming a narrow bound state.

The right panel of fig.~\ref{fig:charmonium} shows $J/\psi$ spectral functions determined using the BR method \cite{Burnier:2013nla}.  These results seem to suggest a substantial positive mass shift and broadening already in the hadronic phase.  However, much of this effect has been found \cite{Kelly:2018hsi} to be an artefact of the limited temporal extent at higher temperature, as spectral functions obtained from reconstructed correlators \cite{Ding:2012sp} exhibit the same features.  A comparison with such spectral functions, as was done in Ref.~\cite{Kelly:2018hsi}, will therefore be required to determine the extent of thermal modifications.

\subsection{Beauty hadrons}
\label{sec:beauty}

\begin{figure*}[h]
\centering
\includegraphics[width=\colw,clip]{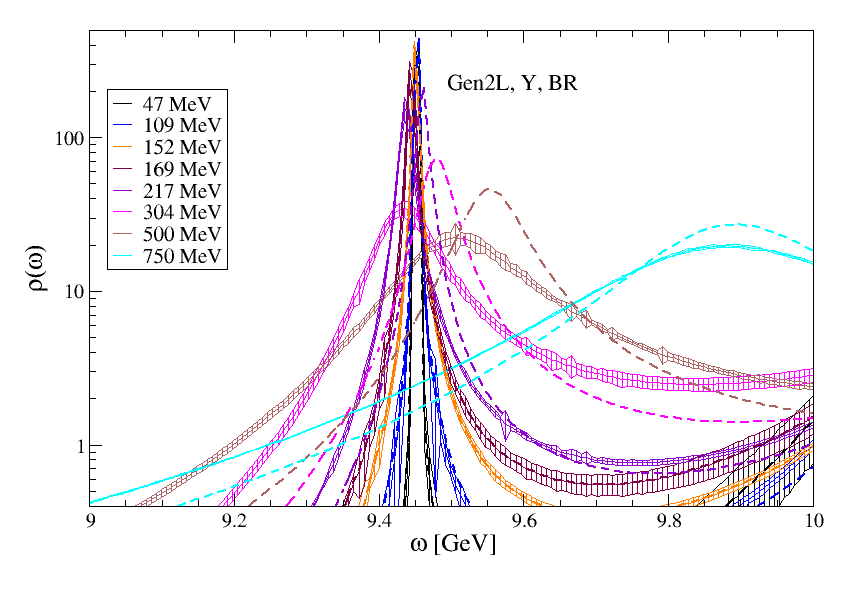}
\includegraphics[width=\colw,clip]{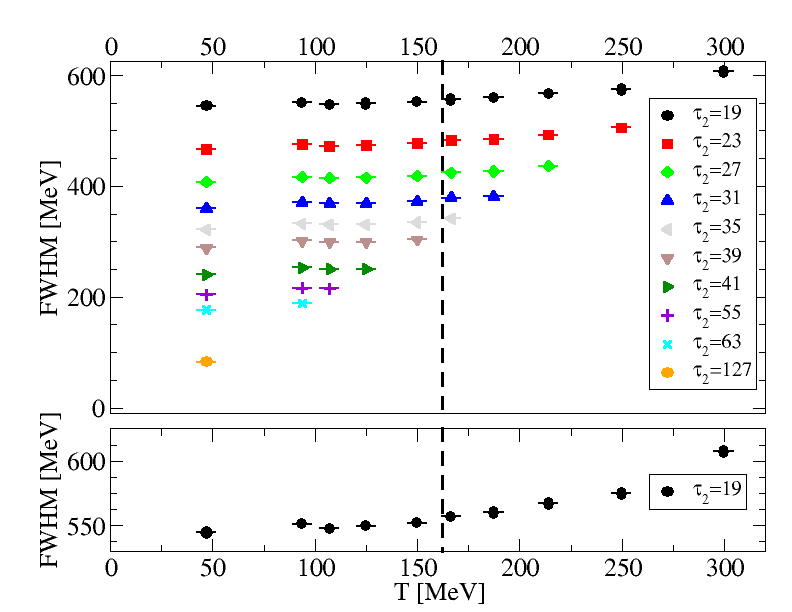}
\caption{Left: $\Upsilon$ spectral functions from the BR method.  The dashed curves denote the spectral functions obtained from truncated correlators at the lowest available temperature, see text for details.  The uncertainty bands on these spectral functions (not shown) are of the same magnitude as those from the $T>0$ correlators.  Right: $\Upsilon$ width from fits using a Gaussian spectral function Ansatz \cite{Spriggs:2021dsb}.  The top panel shows how the fitted width depends on the final point $\tau_2$ of the fit range for each temperature, while the lower panel shows the width as a function of $T$ using the same fit range for all temperatures.} 
\label{fig:beauty}
\end{figure*}

Figure~\ref{fig:beauty} shows selected results for medium modifications of $\Upsilon$ mesons using nonrelativistic QCD (NRQCD).  These are part of an ongoing investigation using a wide range of different methods.  Some results from this investigation have been presented in Ref.~\cite{Spriggs:2021dsb}.  The left hand plot shows spectral functions obtained using the BR method for selected temperatures, compared to the spectral functions $\rho_{\text{rec}}(\omega)$ obtained from correlators at the lowest available temperature, $T=47$, denoted by dashed lines.  Since the NRQCD kernel, $K(\tau;T)=e^{-\om\tau}$, is temperature-independent, the reconstructed correlator may be obtained by simply discarding points with $\tau>1/T$ from the correlator at the reference temperature.  We find that the spectral functions agree perfectly with those from the reconstructed (truncated) correlators for $T\lesssim T_c$.  For $T>T_c$ the two diverge, with $\rho(\om)$ having a peak at lower $\om$ than $\rho_{\text{rec}}(\om)$ as well as a slightly larger width.  This suggests a small, negative mass shift.  A quantitative analysis is in progress.

The right hand panel of fig.~\ref{fig:beauty} shows the $\Upsilon$ width determined from fits to the correlators using a Gaussian Ansatz for the spectral function \cite{Spriggs:2021dsb}.  Both the mass and the width are found to depend strongly on the fit window used, so a comparison of results using the same fit window at different temperatures is mandatory to disentangle real physical effects from this artefact.  Such a comparison is shown in the upper panel.  It was found \cite{Spriggs:2021dsb} that an extrapolation to $\tau_2=\infty$ at the lowest temperature gave a width consistent with zero.  The bottom panel shows the width as function of temperature using the same fit window for all $T$; we see no change below $T_c$ and an increase of up to 60 MeV at $T=300\,$MeV.

\section{Tuning for new ensembles}
\label{sec:tuning}

\begin{figure}[t]
\centering
\includegraphics[width=\colw,clip]{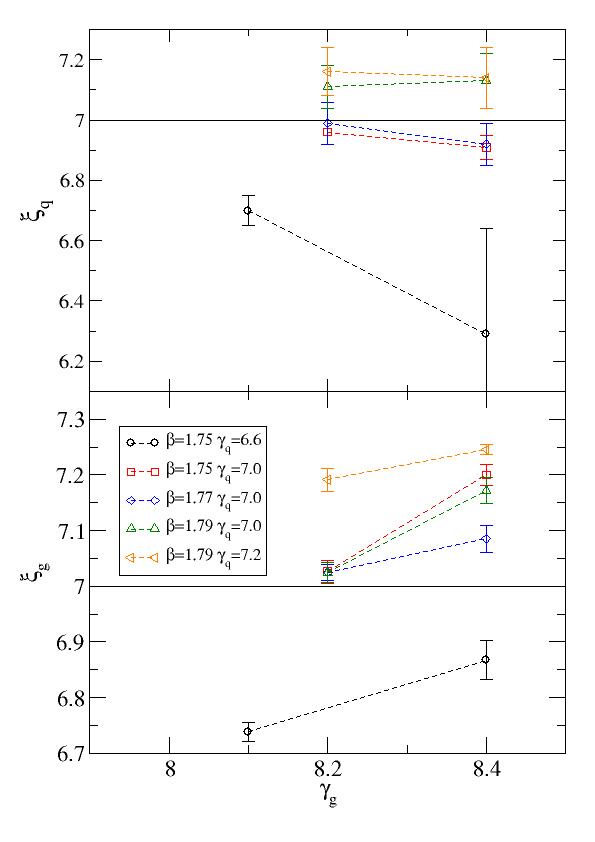}
\includegraphics[width=\colw,clip]{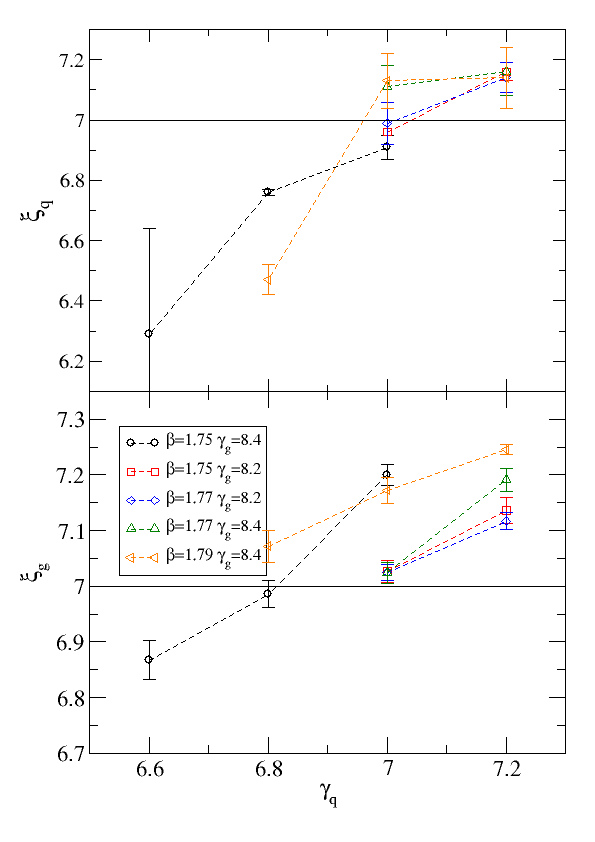}
\caption{Anisotropy tuning for Gen3, showing how the output
  anisotropies $\xi_{g,q}$ depend on the input gauge anisotropy
  $\gamma_g$ (left) and fermion anisotropy $\gamma_q$ (right).}
\label{fig:tuning}
\end{figure}

While analysing data from our Gen2L ensembles, we are preparing for next-generation ensembles which will control the systematic effects of the finite lattice spacing and heavier-than-physical light quark mass.  Specifically, our Gen2P ensemble will bring us to the physical pion mass with a lattice spacing and anisotropy approximately equal to those of Gen2 and Gen2L.  This involves tuning the bare quark mass to reproduce the physical pion mass; this has been done on an $L\approx3\,$fm volume and will be finalised on an $L\approx5\,$fm volume.

The Gen3 ensemble will double the anisotropy and hence halve the temporal lattice spacing while keeping the spatial lattice spacing and pion mass approximately equal to the Gen2 values, $a_s\approx0.12\,$fm and $m_\pi\approx400\,$MeV.  Obtaining the values of the input parameters, the bare gauge and fermion anisotropies $\gamma_g, \gamma_q$, the gauge coupling $\beta$ and the bare quark mass $m_0$, to produce these outputs is a 4-dimensional simultaneous tuning process.  Fig.~\ref{fig:tuning} shows examples of how the physical gauge anisotropy $\xi_g$, determined from the Symanzik flow \cite{Borsanyi:2012zr,Borsanyi:2012zs}, and the fermion anisotropy $\xi_q$, determined from the pion dispersion relation, depend on $\gamma_g$ and $\gamma_q$.  To date we have determined the parameters for the 3-flavour symmetric point with the target value $m_{PS}/m_V=0.545$ \cite{Kelly:PhD} and are in the process of tuning the strange and light quark mass parameters separately.

\section{Summary}

We have presented results for medium modifications, in particular mass shifts, of mesons and baryons containing light, strange, charm and beauty quarks, using anisotropic lattice QCD with 2+1 flavours of clover-improved Wilson quark and $m_\pi\approx240\,$MeV.  We find that the chiral partners $\rho$ and $a_1$ become degenerate at the chiral crossover temperature, with the masses of both decreasing below $T_c$, while the pion mass increases.  For mesons containing charm quarks we find a small negative mass shift below $T_c$.  Singly-charmed baryons exhibit a parity-doubling behaviour similar to that observed for light baryons, while doubly-charmed baryons show a much reduced sensitivity to the chiral crossover.  We report on progress in determining the temperature-dependent mass and width of beautonium states in nonrelativistic QCD.
Finally, we report on the progress of determining the parameters of our next-generation ensembles, which will reach the physical pion mass (Gen2P) and reduce the temporal lattice spacing (Gen3).  These ensembles are expected to come on stream in the coming year.

\section*{Acknowledgments}

This work is supported by the UKRI Science and Technology Facili-
ties Council (STFC) Consolidated Grant No. ST/T000813/1 and by STRONG-2020 ``The strong interaction at the frontier of knowledge: fundamental research
and applications'' which received funding from the European Union’s Horizon
2020 research and innovation programme under grant agreement No 824093.
S. K. is supported by the National Research Foundation of Korea under Grant No. NRF-2021R1A2C1092701 funded by the Korean government (MEST).
M. N. A. acknowledges support from The Royal Society Newton International Fellowship. R. H. D. and E. M. acknowledge support from a Maynooth University SPUR scholarship.  R. H. D. is supported by a Maynooth University John and Pat Hume scholarship.

We are grateful to DiRAC, PRACE, ICHEC and Supercomputing Wales for the use of their computing resources and to the Swansea
Academy for Advanced Computing for support. This work was performed using
the PRACE Marconi-KNL resources hosted by CINECA, Italy and the DiRAC
Extreme Scaling service and Blue Gene Q Shared Petaflop system at the University of Edinburgh operated by the Edinburgh Parallel Computing Centre. The
DiRAC equipment is part of the UK’s National e-Infrastructure and was funded
by UK’s BIS National e-infrastructure capital grant ST/K000411/1, STFC capital grants ST/H008845/1 and ST/R00238X/1, and STFC DiRAC Operations grants ST/K005804/1, ST/K005790/1 and ST/R001006/1.
\bibliography{references}

\begin{thebibliography}{25}

\bibitem{Ding:2012sp}
H.~Ding, A.~Francis, O.~Kaczmarek, F.~Karsch, H.~Satz et~al., Phys.Rev.
  \textbf{D86}, 014509 (2012), \texttt{1204.4945}

\bibitem{Aarts:2007pk}
G.~Aarts, C.~Allton, M.B. Oktay, M.~Peardon, J.I. Skullerud, Phys. Rev.
  \textbf{D76}, 094513 (2007), \texttt{0705.2198}

\bibitem{Kelly:2018hsi}
A.~Kelly, A.~Rothkopf, J.I. Skullerud, Phys. Rev. \textbf{D97}, 114509 (2018),
  \texttt{1802.00667}

\bibitem{Aarts:2010ek}
G.~Aarts, S.~Kim, M.P. Lombardo, M.B. Oktay, S.M. Ryan et~al., Phys.Rev.Lett.
  \textbf{106}, 061602 (2011), \texttt{1010.3725}

\bibitem{Aarts:2011sm}
G.~Aarts, C.~Allton, S.~Kim, M.P. Lombardo, M.B. Oktay et~al., JHEP
  \textbf{1111}, 103 (2011), \texttt{1109.4496}

\bibitem{Aarts:2013kaa}
G.~Aarts, C.~Allton, S.~Kim, M.P. Lombardo, S.M. Ryan, J.I. Skullerud, JHEP
  \textbf{1312}, 064 (2013), \texttt{1310.5467}

\bibitem{Aarts:2014cda}
G.~Aarts, C.~Allton, T.~Harris, S.~Kim, M.P. Lombardo et~al., JHEP
  \textbf{1407}, 097 (2014), \texttt{1402.6210}

\bibitem{Aarts:2015mma}
G.~Aarts, C.~Allton, S.~Hands, B.~J{\"a}ger, C.~Praki, J.I. Skullerud (2015),
  \texttt{1502.03603}

\bibitem{Aarts:2017rrl}
G.~Aarts, C.~Allton, D.~De~Boni, S.~Hands, B.~J{\"a}ger, C.~Praki, J.I.
  Skullerud, JHEP \textbf{06}, 034 (2017), \texttt{1703.09246}

\bibitem{Aarts:2018glk}
G.~Aarts, C.~Allton, D.~De~Boni, B.~J{\"a}ger, Phys. Rev. \textbf{D99}, 074503
  (2019), \texttt{1812.07393}

\bibitem{Amato:2013naa}
A.~Amato, G.~Aarts, C.~Allton, P.~Giudice, S.~Hands, J.I. Skullerud,
  Phys.Rev.Lett. \textbf{111}, 172001 (2013), \texttt{1307.6763}

\bibitem{Aarts:2014nba}
G.~Aarts, C.~Allton, A.~Amato, P.~Giudice, S.~Hands, J.I. Skullerud, JHEP
  \textbf{1502}, 186 (2015), \texttt{1412.6411}

\bibitem{Aarts:2020vyb}
G.~Aarts et~al., Phys. Rev. D \textbf{105}, 034504 (2022), \texttt{2007.04188}

\bibitem{Edwards:2008ja}
R.G. Edwards, B.~Joo, H.W. Lin, Phys. Rev. D \textbf{78}, 054501 (2008),
  \texttt{0803.3960}

\bibitem{Lin:2008pr}
H.W. Lin et~al. (Hadron Spectrum Collaboration), Phys.Rev. \textbf{D79}, 034502
  (2009), \texttt{0810.3588}

\bibitem{Wilson:2015dqa}
D.J. Wilson, R.A. Briceno, J.J. Dudek, R.G. Edwards, C.E. Thomas, Phys. Rev. D
  \textbf{92}, 094502 (2015), \texttt{1507.02599}

\bibitem{Wilson:2019wfr}
D.J. Wilson, R.A. Briceno, J.J. Dudek, R.G. Edwards, C.E. Thomas, Phys. Rev.
  Lett. \textbf{123}, 042002 (2019), \texttt{1904.03188}

\bibitem{Aarts:2022krz}
G.~Aarts, C.~Allton, R.~Bignell, T.J. Burns, S.C. Garc\'\i{}a-Mascaraque,
  S.~Hands, B.~J\"ager, S.~Kim, S.M. Ryan, J.I. Skullerud (2022),
  \texttt{2209.14681}

\bibitem{Bignell:InPrep}
M.N. Anwar, R.~Bignell et~al., \emph{Charmed baryons}, in preparation

\bibitem{Quinn:2019uwq}
R.~Quinn, J.~Glesaaen, A.~Rothkopf, J.I. Skullerud, PoS
  \textbf{Confinement2018}, 272 (2019), \texttt{1903.11006}

\bibitem{Burnier:2013nla}
Y.~Burnier, A.~Rothkopf, Phys. Rev. Lett. \textbf{111}, 182003 (2013),
  \texttt{1307.6106}

\bibitem{Spriggs:2021dsb}
T.~Spriggs et~al., EPJ Web Conf. \textbf{258}, 05011 (2022),
  \texttt{2112.04201}

\bibitem{Borsanyi:2012zr}
S.~Borsanyi, S.~Durr, Z.~Fodor, S.D. Katz, S.~Krieg, T.~Kurth, S.~Mages,
  A.~Schafer, K.K. Szabo (2012), \texttt{1205.0781}

\bibitem{Borsanyi:2012zs}
S.~Borsanyi et~al., JHEP \textbf{09}, 010 (2012), \texttt{1203.4469}

\bibitem{Kelly:PhD}
A.~Kelly, Ph.D. thesis, Maynooth University (2017)

\end{thebibliography}

\end{document}